\RequirePackage{lineno}  

\documentclass[aps,prl,twocolumn,superscriptaddress,showpacs]{revtex4} 

\usepackage{graphicx}

\begin{document}
\newcommand{\fix}[1]{} 
\newcommand{\MET}{\mbox{$\raisebox{.3ex}{$\not$}E_{\rm T}$}}

\preprint{CDF/DOC/TOP/CDFR/8642} 
\preprint{Version 1}          

\title{Precise measurement of the top quark mass in the lepton+jets topology at CDF II}

\affiliation{Institute of Physics, Academia Sinica, Taipei, Taiwan 11529, Republic of China} 
\affiliation{Argonne National Laboratory, Argonne, Illinois 60439} 
\affiliation{Institut de Fisica d'Altes Energies, Universitat Autonoma de Barcelona, E-08193, Bellaterra (Barcelona), Spain} 
\affiliation{Baylor University, Waco, Texas  76798} 
\affiliation{Istituto Nazionale di Fisica Nucleare, University of Bologna, I-40127 Bologna, Italy} 
\affiliation{Brandeis University, Waltham, Massachusetts 02254} 
\affiliation{University of California, Davis, Davis, California  95616} 
\affiliation{University of California, Los Angeles, Los Angeles, California  90024} 
\affiliation{University of California, San Diego, La Jolla, California  92093} 
\affiliation{University of California, Santa Barbara, Santa Barbara, California 93106} 
\affiliation{Instituto de Fisica de Cantabria, CSIC-University of Cantabria, 39005 Santander, Spain} 
\affiliation{Carnegie Mellon University, Pittsburgh, PA  15213} 
\affiliation{Enrico Fermi Institute, University of Chicago, Chicago, Illinois 60637} 
\affiliation{Comenius University, 842 48 Bratislava, Slovakia; Institute of Experimental Physics, 040 01 Kosice, Slovakia} 
\affiliation{Joint Institute for Nuclear Research, RU-141980 Dubna, Russia} 
\affiliation{Duke University, Durham, North Carolina  27708} 
\affiliation{Fermi National Accelerator Laboratory, Batavia, Illinois 60510} 
\affiliation{University of Florida, Gainesville, Florida  32611} 
\affiliation{Laboratori Nazionali di Frascati, Istituto Nazionale di Fisica Nucleare, I-00044 Frascati, Italy} 
\affiliation{University of Geneva, CH-1211 Geneva 4, Switzerland} 
\affiliation{Glasgow University, Glasgow G12 8QQ, United Kingdom} 
\affiliation{Harvard University, Cambridge, Massachusetts 02138} 
\affiliation{Division of High Energy Physics, Department of Physics, University of Helsinki and Helsinki Institute of Physics, FIN-00014, Helsinki, Finland} 
\affiliation{University of Illinois, Urbana, Illinois 61801} 
\affiliation{The Johns Hopkins University, Baltimore, Maryland 21218} 
\affiliation{Institut f\"{u}r Experimentelle Kernphysik, Universit\"{a}t Karlsruhe, 76128 Karlsruhe, Germany} 
\affiliation{High Energy Accelerator Research Organization (KEK), Tsukuba, Ibaraki 305, Japan} 
\affiliation{Center for High Energy Physics: Kyungpook National University, Taegu 702-701, Korea; Seoul National University, Seoul 151-742, Korea; SungKyunKwan University, Suwon 440-746, Korea} 
\affiliation{Ernest Orlando Lawrence Berkeley National Laboratory, Berkeley, California 94720} 
\affiliation{University of Liverpool, Liverpool L69 7ZE, United Kingdom} 
\affiliation{University College London, London WC1E 6BT, United Kingdom} 
\affiliation{Centro de Investigaciones Energeticas Medioambientales y Tecnologicas, E-28040 Madrid, Spain} 
\affiliation{Massachusetts Institute of Technology, Cambridge, Massachusetts  02139} 
\affiliation{Institute of Particle Physics: McGill University, Montr\'{e}al, Canada H3A~2T8; and University of Toronto, Toronto, Canada M5S~1A7} 
\affiliation{University of Michigan, Ann Arbor, Michigan 48109} 
\affiliation{Michigan State University, East Lansing, Michigan  48824} 
\affiliation{University of New Mexico, Albuquerque, New Mexico 87131} 
\affiliation{Northwestern University, Evanston, Illinois  60208} 
\affiliation{The Ohio State University, Columbus, Ohio  43210} 
\affiliation{Okayama University, Okayama 700-8530, Japan} 
\affiliation{Osaka City University, Osaka 588, Japan} 
\affiliation{University of Oxford, Oxford OX1 3RH, United Kingdom} 
\affiliation{University of Padova, Istituto Nazionale di Fisica Nucleare, Sezione di Padova-Trento, I-35131 Padova, Italy} 
\affiliation{LPNHE, Universite Pierre et Marie Curie/IN2P3-CNRS, UMR7585, Paris, F-75252 France} 
\affiliation{University of Pennsylvania, Philadelphia, Pennsylvania 19104} 
\affiliation{Istituto Nazionale di Fisica Nucleare Pisa, Universities of Pisa, Siena and Scuola Normale Superiore, I-56127 Pisa, Italy} 
\affiliation{University of Pittsburgh, Pittsburgh, Pennsylvania 15260} 
\affiliation{Purdue University, West Lafayette, Indiana 47907} 
\affiliation{University of Rochester, Rochester, New York 14627} 
\affiliation{The Rockefeller University, New York, New York 10021} 
\affiliation{Istituto Nazionale di Fisica Nucleare, Sezione di Roma 1, University of Rome ``La Sapienza," I-00185 Roma, Italy} 
\affiliation{Rutgers University, Piscataway, New Jersey 08855} 
\affiliation{Texas A\&M University, College Station, Texas 77843} 
\affiliation{Istituto Nazionale di Fisica Nucleare, University of Trieste/\ Udine, Italy} 
\affiliation{University of Tsukuba, Tsukuba, Ibaraki 305, Japan} 
\affiliation{Tufts University, Medford, Massachusetts 02155} 
\affiliation{Waseda University, Tokyo 169, Japan} 
\affiliation{Wayne State University, Detroit, Michigan  48201} 
\affiliation{University of Wisconsin, Madison, Wisconsin 53706} 
\affiliation{Yale University, New Haven, Connecticut 06520} 
\author{T.~Aaltonen}
\affiliation{Division of High Energy Physics, Department of Physics, University of Helsinki and Helsinki Institute of Physics, FIN-00014, Helsinki, Finland}
\author{A.~Abulencia}
\affiliation{University of Illinois, Urbana, Illinois 61801}
\author{J.~Adelman}
\affiliation{Enrico Fermi Institute, University of Chicago, Chicago, Illinois 60637}
\author{T.~Affolder}
\affiliation{University of California, Santa Barbara, Santa Barbara, California 93106}
\author{T.~Akimoto}
\affiliation{University of Tsukuba, Tsukuba, Ibaraki 305, Japan}
\author{M.G.~Albrow}
\affiliation{Fermi National Accelerator Laboratory, Batavia, Illinois 60510}
\author{S.~Amerio}
\affiliation{University of Padova, Istituto Nazionale di Fisica Nucleare, Sezione di Padova-Trento, I-35131 Padova, Italy}
\author{D.~Amidei}
\affiliation{University of Michigan, Ann Arbor, Michigan 48109}
\author{A.~Anastassov}
\affiliation{Rutgers University, Piscataway, New Jersey 08855}
\author{K.~Anikeev}
\affiliation{Fermi National Accelerator Laboratory, Batavia, Illinois 60510}
% ...
\collaboration{CDF Collaboration\footnote{With visitors from $^a$University of Athens, 15784 Athens, Greece, 
$^b$University of Bristol, Bristol BS8 1TL, United Kingdom, 
$^c$University Libre de Bruxelles, B-1050 Brussels, Belgium, 
$^d$Cornell University, Ithaca, NY  14853, 
$^e$University of Cyprus, Nicosia CY-1678, Cyprus, 
$^f$University College Dublin, Dublin 4, Ireland, 
$^g$University of Edinburgh, Edinburgh EH9 3JZ, United Kingdom, 
$^h$University of Heidelberg, D-69120 Heidelberg, Germany, 
$^i$Universidad Iberoamericana, Mexico D.F., Mexico, 
$^j$University of Manchester, Manchester M13 9PL, England, 
$^k$Nagasaki Institute of Applied Science, Nagasaki, Japan, 
$^l$University de Oviedo, E-33007 Oviedo, Spain, 
$^m$University of London, Queen Mary College, London, E1 4NS, England, 
$^n$University of California Santa Cruz, Santa Cruz, CA  95064, 
$^o$Texas Tech University, Lubbock, TX  79409, 
$^p$University of California, Irvine, Irvine, CA  92697, 
$^q$IFIC(CSIC-Universitat de Valencia), 46071 Valencia, Spain, 
}}
\noaffiliation



\date{\today}

\begin{abstract}

We present a measurement of the mass of the top quark from
proton-antiproton collisions recorded at the CDF experiment in Run II
of the Fermilab Tevatron. We analyze events from the single lepton
plus jets final state ($t\bar t\rightarrow W^+bW^-\bar b\rightarrow
l\nu b q \bar q'\bar b$). The top quark mass is extracted using a
direct calculation of the probability density that each event
corresponds to the $t\bar t$ final state. The probability is a
function of both the mass of the top quark and the energy scale of the
calorimeter jets, which is constrained {\it in situ}
by the hadronic $W$ boson mass. Using 167 events observed
in 955~pb$^{\mbox{-}1}$ of integrated luminosity, we achieve the
single most precise measurement of the top quark mass, 170.8 $\pm$ 2.2
(stat.) $\pm$ 1.4 (syst.)  GeV/$c^2$.

\end{abstract}

\pacs{14.65.Ha, 13.85Qk, 12.15Ff}

\maketitle



The top quark is the heaviest known elementary particle, about 40
times more massive than the next-heaviest fermion. Its large mass plays an
important role in loop corrections to several electroweak observables
predicted by the standard model. In conjunction with measurements of
the $W$ boson mass, a precise measurement of the top quark mass,
$m_t$, constrains the mass of the as yet unobserved Higgs boson
\cite{theory}. Moreover, by comparing precision
electroweak measurements to predictions including the relevant loop
corrections, a precise measurement of $m_t$ can help constrain
contributions from physics beyond the standard model. Measuring $m_t$
to the highest achievable precision is therefore one of the main goals
of the experiments operating at the Fermilab Tevatron collider.
 
At the Tevatron, top quarks are mainly produced in pairs.  They decay
essentially 100$\%$ of the time into a $W$ boson and a $b$ quark, with
the $W$ decaying into quarks or leptons. The result presented here
uses the lepton+jets channel, where one $W$ decays into two quarks,
and the other decays to an electron or a muon and the corresponding
neutrino. In the past, this channel has provided the most precise
measurements of $m_t$, and recent measurements can be found in
Ref.~\cite{topmass}.

In this letter we report the single most precise measurement of the
top quark mass from Tevatron proton-antiproton collisions at
$\sqrt{s}$=1.96 TeV, using 955~pb$^{\mbox{-}1}$ of integrated
luminosity collected with the CDF II detector from February 2002 to
March 2006.
The CDF II detector is a general-purpose particle detector and is
described in detail elsewhere~\cite{CDF}. It has a solenoidal charged
particle spectrometer, consisting of 7-8 layers of silicon microstrip
detectors and a cylindrical drift chamber immersed in a 1.4~T magnetic
field, a segmented sampling calorimeter, and a set of charged particle
detectors outside the calorimeter used to identify muon candidates. We
use a right-handed cylindrical coordinate system with the origin in
the center of the detector, where $\theta$ and $\phi$ are the polar
and azimuthal angles, respectively, and pseudorapidity is defined as
$\eta\equiv - \ln$ $\rm tan(\theta/2)$.  Transverse energy and
momentum are $E_T$=$E\sin\theta$ and $p_T$=$p\sin\theta$,
respectively, where $E$ and $p$ are energy and momentum.

Events in the lepton+jets decay channel are selected to have a single,
isolated electron or muon candidate with large transverse energy,
large imbalance in transverse momentum in the event (missing
transverse energy~\cite{met}, \MET) as expected from the undetectable
neutrino, and exactly four jets with large transverse energy. Jets are
reconstructed using a cone algorithm with radius $\Delta R
\equiv\sqrt{(\Delta \eta)^2+(\Delta \phi)^2} = 0.4$. Of these jets, we
require at least one to have originated from a $b$ quark by using an
algorithm that identifies a long-lived $B$ hadron through the presence
of a displaced vertex ($b$ tag)~\cite{secvtx}.  Backgrounds to the
$t\bar t$ signal arise from multi-jet QCD production (non-$W$), $W$
production in association with jets ($W$+jets), and electroweak
backgrounds (EWK) composed of diboson ($WW$, $WZ$, $ZZ$) and single
top production.  $W$+jets background events include jets with real $b$
flavor as well as light flavor jets incorrectly identified as $b$
jets. To remove the non-$W$ backgrounds, where \MET~is due to
mis-measured jet energies, we require \MET~not to be aligned with the
highest energy jet by a suitable requirement on $\Delta\phi$ between
this jet and~\MET. Table~\ref{tab:cuts} summarizes the selection
criteria used in this analysis, and a more detailed description can be
found in Ref.~\cite{xsection}. We select 167 events of which we expect
about 85$\%$ to be $t\bar t$ events.  Table~\ref{tab:events} shows the
expected sample composition determined with
318~pb$^{\mbox{-}1}$~\cite{xsection}, scaled to 955~pb$^{\mbox{-}1}$
and assuming a $t\bar t$ cross section of 8.0~pb. Detailed background
estimate studies justify this scaling, and residual differences are
absorbed in the systematic background uncertainty.
 
\begin{table}[h]
\caption{Event selection criteria~\cite{xsection}. }
\begin{center}
\begin{tabular}{lc}
\hline
\hline
Lepton & $E_T>$20 GeV (electron, muon), $|\eta| < 1$ \\ Jets & exactly
4 with $E_T > 15$ GeV, $|\eta| < 2.0$ \\
\MET                     &  $> 20$ GeV, calculated over $|\eta| < 3.6$ \\
$b$ tag jets             &  $\ge 1$ from a secondary vertex, $|\eta| < 1.5$\\
Non-$W$ veto             &  $0.5 \le \Delta\phi \le 2.5$ when $\MET < 30$ GeV \\
\hline
\hline
\end{tabular}
\end{center}
\label{tab:cuts}
\end{table}

\begin{table}[h]
\caption{Background composition and expected number of $t\bar t$ candidates. All uncertainties are statistical only.}
\begin{center}
\begin{tabular}{lc}
\hline
\hline
Source & Expected number of events \\
\hline
$W$+jets     &  14.5 $\pm$ 5.1 \\
non-$W$      &  5.2  $\pm$ 2.6 \\
EWK          &  2.2  $\pm$ 0.5 \\
\hline
Total        & 22.0  $\pm$ 8.2 \\
\hline
$t\bar t$ ($\sigma$=8.0 pb, $m_t$=170 GeV/c$^2$)         &  145.1 $\pm$ 16.5 \\
\hline
\hline
Data        & 167 \\
\hline
\hline
\end{tabular}
\end{center}
\label{tab:events}
\end{table}

We analyze the selected events using a likelihood technique that
relies on calculations of probability densities based on matrix
elements for the signal ($t\bar{t}$) and dominant background
($W$+jets) processes~\cite{ME}. The backgrounds other than
$W$+jets are found to be adequately described by the $W$+jets probability
density. Given a set of observed variables, ${x}$, and underlying
partonic quantities, $y$, the signal and background probability
densities are constructed by integrating over the appropriate
parton-level differential cross section, $d\sigma(y)/dy$, convolved
with parton distribution functions (PDFs) and detector resolution
effects:
\begin{equation}
P({x})  = \sum_{jet \ perm.} \int  \frac{d\sigma(y)}{dy} f(q_1) f(q_2) dq_1 dq_2 W({x},{y})dy. 
\label{eq:prob}
\end{equation}

\noindent The PDFs ($f(q_1)$ and $f(q_2)$) take into account the flavors 
of colliding quark and antiquark and are given by CTEQ5L
\cite{CTEQ}. The detector resolution effects are described by a
transfer function $W({x},{y})$ relating $x$ to $y$. The momenta of the
leptons and the angles of jets and leptons are taken to be exactly
measured, and therefore $W({x},{y})$ for these quantities is given by
the product of Dirac delta functions. The non-trivial part of
$W({x},{y})$ maps parton energies to measured jet energies after
correction for instrumental detector effects~\cite{jesnim}. This
mapping is obtained by parameterizing the jet response in fully
simulated $t\bar t$ events created by the Monte Carlo (MC) generator
{\sc pythia}~\cite{MC}, and including the effects of radiation,
hadronization, measurement resolution, and energy omitted from the jet
cone by the reconstruction algorithm. The $t\bar t$ and $W$+jets
probability densities, $P_{t\bar t}$ and $P_{W+jets}$, include all
possible permutations of matching jets with partons as well as all
possible longitudinal momenta for the neutrino in the $W$ decay. The
permutations are reduced to six or two by exploiting $b$ tagging
information (single-tag or double-tag, respectively). We use different
transfer functions for light quark jets and $b$ jets, depending on the
flavor of the parton assigned to the jet. In calculating
$d\sigma(y)/dy$, $P_{t\bar t}$ uses the leading order matrix element
of the $q\bar q\rightarrow t\bar t$ process~\cite{mahlonparke}, and
$P_{W+jets}$ uses the sum of matrix elements of the $W$+4 jets
subroutines encoded in the {\sc vecbos} Monte Carlo generator
\cite{vecbos}.

The final state described by $d\sigma(y)/dy$ contains 6 particles,
which introduces 20 integration variables in Eq.~\ref{eq:prob},
including the longitudinal momenta of the incoming quarks. By imposing
energy-momentum conservation, in conjunction with the Dirac delta
functions in $W({x},{y})$ we reduce the dimensionality of the
remaining integration to five.  The integration in $P_{W+jets}$ is
performed over the energies of the outgoing partons and the invariant
mass of the leptonically decaying $W$ using a Monte Carlo
technique. In order to reduce the calculation time for $P_{t\bar t}$,
we integrate over the following variables: the invariant masses of
$t$, $\bar t$, $W^+$, and $W^-$, and the energy of one of the quarks
from the hadronic $W$ decay. Our method includes two additional
integrations over the transverse momentum components of the $t\bar{t}$
system. The integration in $P_{t\bar t}$ uses the numerical
integration code {\sc vegas}~\cite{vegas}.

The largest potential systematic uncertainty in this measurement
arises from the energy scale of jets. To decrease this uncertainty, we
exploit the fact that the hadronically decaying $W$ provides an {\it
in situ} constraint of the jet energy scale, as the two jets should
form an invariant mass consistent with the precisely known mass of the
$W$ boson \cite{theory}. The jet energy scale and the mass of the top
quark are simultaneously determined from a two-dimensional likelihood
that includes their correlation. A salient feature of this method is
that the uncertainty due to the jet energy scale will be reduced with
increasing statistics. Thus $P_{t\bar t}$ is evaluated as a function
of $m_{t}$ and an assumed jet energy scale factor $f_{JES}=
E_{jet}^{obs}/E_{jet}$, where $E_{jet}^{obs}$ is the observed jet
energy and $E_{jet}$ is the true jet energy.

To extract $m_t$ and $f_{JES}$ from the data, we build a likelihood
function for $N$ selected events by adding $P_{t\bar t}$ and
$P_{W+jets}$ for each event.  The combined likelihood is minimized
with respect to three variables: $m_{t}$, $f_{JES}$, and $C_s$, the
fraction of events consistent with our $t\bar t$ signal hypothesis.
The likelihood for $N$ events is given by
\begin{eqnarray}
& & \mathcal{L}(x_1, x_2, ..., x_N;m_{t},f_{JES},C_s)  =  \nonumber \\
& & e^{-N (C_s \langle A_{t\bar{t}}(m_{t},f_{JES}) \rangle+(1-C_s) \langle A_{W+\mathrm{jets}}(f_{JES}) \rangle )} \times \nonumber \\
& & \prod_{i=1}^N [C_s P_{t\bar{t}}(x;m_{t},f_{JES}) + (1 - C_s)
    P_{W+\mathrm{jets}}(x;f_{JES})] \nonumber \\
\label{eqn:finallike}
\end{eqnarray}
\noindent where the first factor arises from the Poisson extension of
the likelihood and normalizes the combined event probability density,
and $\langle A \rangle$ refers to the mean acceptance for $t\bar t$ or
$W$+jets events. We use fully simulated MC $t\bar t$ and $W$+jets
events to determine the functional form of $\langle A \rangle$.
$P_{W+jets}$ is evaluated at the central jet energy scale factor,
$f_{JES}$=1.  The $f_{JES}$ dependence of $P_{W+jets}$ is determined
by varying the input $f_{JES}$ in MC event samples ($f_{JES}^{MC}$)
and by parameterizing the average likelihood response as a function of
$f_{JES}$. We use the $m_t$ dependence of the theoretical leading
order $t\bar{t}$ cross section to normalize $P_{t\bar t}$. 
Because we use a leading order matrix element to calculate $P_{t\bar
t}$, we find that $t\bar t$ events where at least one of the four
reconstructed jets cannot be matched to a parton from the $t\bar t$
decay within $\Delta R<0.4$ behave like background events. As a
consequence, a pure sample of $t\bar t$ events yields $C_s$ of 
0.8. The quoted $C_s$ values are corrected for this effect.
For each event $P_{t\bar t}$ is evaluated in increments
of 2 GeV/$c^2$ in $m_t$ and 0.02 in $f_{JES}$.  At each point of this
grid we fit the entire sample of $N$ events according to 
Eq. \ref{eqn:finallike} and 
the most likely value of $C_s$ is determined using {\sc
minuit}~\cite{MINUIT}. The optimal parameters $m_t$ and $f_{JES}$ are
obtained by fitting the likelihood using a two-dimensional
Gaussian. The statistical uncertainty on $m_t$ includes the
uncertainty on $f_{JES}$.

\begin{figure}[h]
\includegraphics[width=0.47\textwidth]{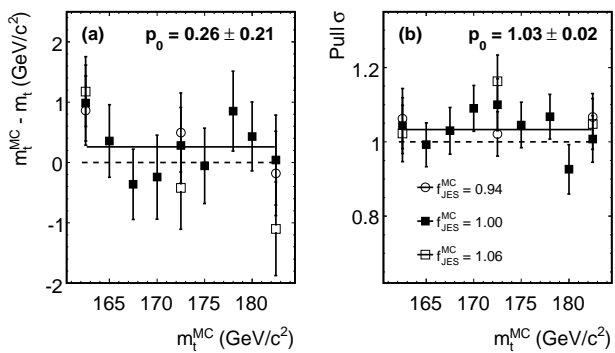}
\caption{Results of pseudo-experiment tests. (a) Difference between the measured $m_t$ and the input top quark mass in the MC event sample ($m_t^{MC}$), as a function of $m_t^{MC}$. (b) Gaussian $\sigma$ of pull distributions (see text), as a function of $m_t^{MC}$. The plots include results using MC event samples with different values of $f_{JES}$ ($f_{JES}^{MC}$). The weighted average $p_0$ is indicated by the solid horizontal line. The dashed line indicates an example of an unbiased result.}
\label{fig:tests}
\end{figure}

The performance of the analysis is tested by extracting $m_t$ from
MC pseudo-experiments containing $t\bar t$ signal samples with various
input top quark masses ($m_t^{MC}$) and background samples 
described in Table II.  The
signal and electroweak background samples are generated using {\sc
herwig}~\cite{MC}. The $W$+jets background is generated using {\sc
alpgen}~\cite{MC} with hadronization and fragmentation done by {\sc
herwig}.  The non-$W$ background is extracted from an independent data
sample. All of the MC samples are processed by the CDF detector
simulation. We construct pseudo-experiments of signal and background
events by fluctuating the number of events around the values shown in
Table~\ref{tab:events}.  Figure
\ref{fig:tests}(a) shows that the fitted Gaussian mean $m_t$ extracted 
from 200 pseudo-experiments per point is unbiased with respect to
$m_t^{MC}$ up to the statistical uncertainty of 0.21~GeV/$c^2$, which
is taken as a systematic uncertainty. Similar tests are performed for
the output of $f_{JES}$. In this case, we find that a bias of +4$\%$
in $f_{JES}$ is present, independent of $m_t$. We correct for this
bias to properly interpret the output of
$f_{JES}$. Figure~\ref{fig:tests}(b) shows the top mass pull width,
defined as the Gaussian $\sigma$ of the top mass residual (${m_t-m_t^{MC}}$)
divided by the uncertainty in each pseudo-experiment $\sigma_{m_t}$,
as a function of $m_t^{MC}$. The pull width is 3\%$\pm$2\% larger than
one on average, and thus, the statistical uncertainty is scaled up by
3\%.


\begin{figure}[h]
\includegraphics[width=0.45\textwidth]{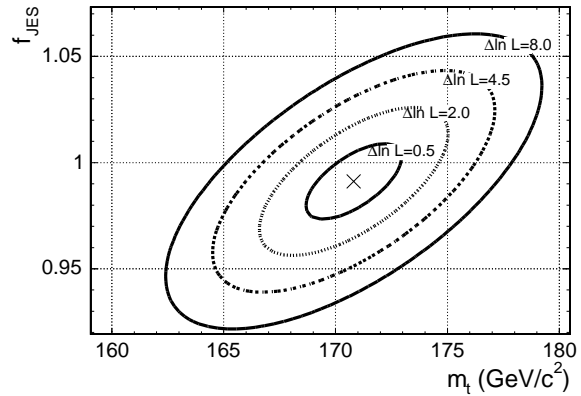}
\caption{Contours of likelihood evaluated over the 955~pb$^{\mbox{-}1}$ event sample. Our measurement is indicated by the $X$.}\label{fig:data}
\end{figure}

\begin{figure}[h]
\includegraphics[width=0.47\textwidth]{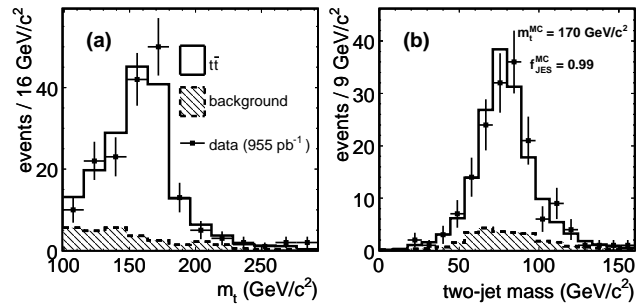}
\caption{Comparison of two kinematic variables for data and MC using
$f_{JES}^{MC}$=0.99 and $m_t^{MC}$=170 GeV/c$^2$. (a) Most probable
value of $m_t$ for each event extracted from evaluating $P_{t\bar t}$
at $f_{JES}$=1. (b) Invariant mass of the pair of jets assigned as $W$
decay products calculated using the most probable permutation at the
most probable value of $m_t$ and $f_{JES}$ in each event evaluated
from $P_{t\bar t}$. The backgrounds contain all contributions shown in
Table~\ref{tab:events}.}\label{fig:mtop}
\end{figure}

Applying this method to data, we measure the top quark
mass to be $m_t$=170.8$\pm$2.2(stat.), and the $f_{JES}$ scale to be
$f_{JES}$=0.99$\pm$0.02(stat.) in good agreement with the reference scale
from the default CDF calibration~\cite{jesnim}. Figure
\ref{fig:data} shows the fitted two-dimensional likelihood with 
 $\Delta\ln L$ contours. The statistical uncertainty is taken from the
maximum and minimum $m_t$ values on the $\Delta\ln L$=0.5 contour. We
find a correlation coefficient of 0.32 between $m_{t}$ and
$f_{JES}$. The fit yields a signal fraction
$C_s$=0.84$\pm$0.10(stat.), which corresponds to 140$\pm$17 $t\bar t$
events and is consistent with the expectation shown in
Table~\ref{tab:events}. Monte Carlo tests have shown that the
resulting $m_t$ is stable over a wide range of sample
purities. Figure~\ref{fig:mtop} shows comparisons of two
representative kinematic quantities between data and simulation using
$f_{JES}^{MC}$=0.99 and $m_t^{MC}$=170 GeV/c$^2$.

The sources of systematic uncertainty are listed in Table
\ref{tab:syssum}. To first order, $f_{JES}$ is already included in the
statistical uncertainty, but we also consider a dependence of
$f_{JES}$ on the $p_T$ and $\eta$ of the jets (residual jet energy
scale) using the dependence found in other studies
\cite{jesnim}. Another uncertainty is calculated from possible
differences in $f_{JES}$ between $b$ jets and light quark jets.  The
generator uncertainty takes into account differences in parton
showering and jet fragmentation between two different MC programs used
to generate $t\bar t$ events, {\sc pythia} and {\sc herwig}.
Variations in initial- and final-state radiation (ISR, FSR),
constrained by studies using Drell-Yan data, are also
considered~\cite{topmass}.  PDF uncertainties are evaluated using MC
samples generated with {\sc mrst}~\cite{mrst} and the full set of
eigenvectors from {\sc cteq6m}~\cite{CTEQ}.  Systematic effects on the
modeling of the background samples include fluctuations in the total
background contribution, relative contributions from individual
background processes, and variations due to the $Q^2$ scale used in
$W$+jets simulation. We also include effects from the uncertainties in
the simulated lepton $p_{T}$, the dependence of $b$ tagging with jet
$p_T$, and the effect of the limited MC event samples used in the
analysis.  Finally, we include possible mis-modeling of multiple
interactions in the simulation at high luminosity. The sum in
quadrature of all systematic uncertainties is 1.4~GeV/$c^2$.

\begin{table}[h]
\caption{Summary of systematic uncertainties.}
\begin{center}
\begin{tabular}{lc}
\hline
\hline
Source & Uncertainty (GeV/$c^2 $) \\
\hline
Residual jet energy scale                   &  0.4  \\
$b$ jet energy scale                        &  0.6  \\
Generator                                   &  0.2  \\
ISR/FSR                                     &  1.1  \\
PDFs                                        &  0.1  \\
Background                                  &  0.2  \\
Lepton $p_{T}$                              &  0.2  \\
$b$ tag $p_T$ dependence                    &  0.3  \\
Monte Carlo statistics                      &  0.2  \\
Multiple interactions                       &  0.1 \\
\hline
Total                                       &  1.4  \\
\hline
\hline
\end{tabular}
\end{center}
\label{tab:syssum}
\end{table}

In summary, we present a measurement of the top quark mass in the
lepton+jets channel using 955~pb$^{\mbox{-}1}$ of data collected by
the CDF experiment.  A matrix element analysis was used with an $in$
$situ$ measurement of the jet energy scale. We measure
\begin{equation}
m_{t} = 170.8 \pm 2.2 \ {\rm (stat.)} \pm 1.4 \ {\rm (syst.)} \ \ {\rm
 GeV/}c^2
\end{equation}
where the statistical uncertainty includes the uncertainty of 1.5
GeV/c$^2$ due to the jet energy scale.  With a total uncertainty of
1.5\%, this result is the most precise measurement of the top quark
mass to date and is a 35\% improvement over the previous best
measurement~\cite{topmass}.

\begin{acknowledgments}

We thank the Fermilab staff and the technical staffs of the
participating institutions for their vital contributions. This work
was supported by the U.S. Department of Energy and National Science
Foundation; the Italian Istituto Nazionale di Fisica Nucleare; the
Ministry of Education, Culture, Sports, Science and Technology of
Japan; the Natural Sciences and Engineering Research Council of
Canada; the National Science Council of the Republic of China; the
Swiss National Science Foundation; the A.P. Sloan Foundation; the
Bundesministerium f\"ur Bildung und Forschung, Germany; the Korean
Science and Engineering Foundation and the Korean Research Foundation;
the Science and Technology Facilities Council and the Royal Society,
UK; the Institut National de Physique Nucleaire et Physique des
Particules/CNRS; the Russian Foundation for Basic Research; the
Comisi\'on Interministerial de Ciencia y Tecnolog\'{\i}a, Spain; the
European Community's Human Potential Programme; the Slovak R\&D Agency;
and the Academy of Finland.

\end{acknowledgments}

\bibliography{topmassprlreview.bib}

\end{document}